\documentclass[backend=biber,preprint,12pt]{elsarticle}

\newcounter{bla}

\usepackage{amsmath}
\journal{Computer Physics Communications}

\begin{document} 
\begin{frontmatter}

\title{ Quark Contraction Tool - QCT}
\author[a]{D.~Djukanovic} 
\address[a]{Helmholtz Institute Mainz, University of Mainz, D-55099 Mainz,
Germany} 
\begin{abstract} 
We present a Mathematica package for the calculation of Wick contractions in
quantum field theories - QCT. The package aims at automatically
generating code for the calculation of physical matrix elements, suitable for
numerical evaluation in a C++ program. To that end commonly used algebraic
manipulations for the calculation of matrix elements in lattice QCD
are implemented.
\end{abstract} 
\end{frontmatter}

{\bf PROGRAM SUMMARY}

\begin{small}
\noindent
{\em Program Title:}  QCT                                        \\
{\em Licensing provisions:} GPLv3                                   \\
{\em Programming language:} Mathematica                                   \\
{\em Nature of problem:}\\
Wick contractions of matrix elements in quantum field theories and in particular in Lattice quantum chromodynamics (Lattice QCD).
\\
{\em Solution method:}\\
The implementation is based on symbolic manipulations of non commuting objects
in Mathematica. The results can be expressed in various formats, including C++
which allows for fast implementation of correlation functions.
\\
{\em Additional comments including Restrictions and Unusual features:}\\
Currently the contractions assume Grassmann valued fields. The program can visualize the contraction in the form of directed graphs.
   \\

\end{small}

\section{Introduction} 
Matrix elements in lattice QCD are typically
expressed as expectation values of time ordered products of quark operators,
which can be computed by means of Wick contractions \cite{Wick:1950ee}.
The aim of this package is to simplify and automate this straightforward but
error prone task. \\
Currently only fermionic contractions are supported, i.e. the fields are
assumed to be Grassmann valued. The complexity of the calculation grows as $N!$,
where $N$ is the number of creation or annihilation operators in the matrix
element, i.e. the total number of quark field operators is $2N$. The
contraction of 16 quark fields of identical type takes roughly 30
seconds\footnote{Timings refer to a system with an Intel Core i5 running @ 2.6
GHz.} generating 40320 terms.  
\section{Basic functionality} 
The package provides functions to perform Wick contractions and
subsequently simplify the expressions, i.e. contract indices and rewrite the
results into QDP++ \cite{Edwards:2004sx} mnemonics. For the Wick contractions the package expects
fermionic field operators, where the fields have color and spin indices, and
depend on the position. The creation (annihilation) operators are  denoted  by
\begin{verbatim}
  Field{B}[<type>,<color index>,<spin index>, <position>]
\end{verbatim}

The simplest contraction thus reads
\begin{eqnarray}
	f^a_\mu(x)\bar{f}^b_\nu(y)=S^{f,ab}_{\mu\nu}(y,x)
\end{eqnarray}
where the fermion propagator $S$ of particles of type $f$ connects point y to
x.  This contraction can be done within the package as
\begin{verbatim} 
In[1]:=
  WickContract[Field[f,a,mu,x]**FieldB[f,b,nu,y]]
Out[1]=
  DE[{f, f}, {y, x}][CI[{a, b}], SI[{mu, nu}]]
\end{verbatim} 
where $**$ is the builtin NonCommutativeMultiply operator, and {\tt DE} denotes
the propagator of a particle of type f with the color (spin) indices enclosed
in the function {\tt CI (SI)}. The package defines formating rules for certain
expressions, i.e. in Mathematica's TraditionalForm or TeXForm the output is
specially typeset, e.g. the color indices appear in the exponent and the spin
indices are displayed as subscripts. The above expression in TeXForm reads
\begin{verbatim}
In[2]:=
  TeXForm[DE[{f, f}, {y, x}][CI[{a, b}], SI[{mu, nu}]]]
Out[2]=
  S_{\mu \nu }^{f,ab}(y,x)
\end{verbatim}
There is a convenience function {\tt ToQDP}, which rewrites the expressions in
a format suitable for calculation within the  QDP++ \cite{Edwards:2004sx}
package. For the above term the conversion reads
\begin{verbatim}
In[3]:=
  ToQDP[%%]
Out[3] =
  /* 
  Result is of type SpinColorMatrix
  {S^f1(x1,y1) -> quarkProp1}
  */ 
  quarkProp1
\end{verbatim}
where in the comment section a description of the QDP++ type that is returned is
given, together with a list of abbreviations for propagator objects. The
abbreviations are not \emph{canonical}, i.e. different expressions might
give different abbreviations, however the package keeps a lookup table for
abbreviations of the quark propagators within a Mathematica session.
Additionally one can provide a list of
replacement rules as a second argument, e.g. special characters or
sub-/superscripted quantities where the CForm would generate undesired terms.
Furthermore the package provides the function {\tt QuarkContract}, which
performs a variety of manipulations to arrive at simpler forms, e.g. 
index summations
\begin{verbatim}
In[4] :=
  QuarkContract[WickContract[
  Pol[SI[{nu, mu}]] ** Field[f, a, mu, x] ** FieldB[f, b, nu, y]
  ]]
Out[4] =
  traceSpin[Pol . DE[{f, f}, {y, x}]][CI[{a, b}]]
\end{verbatim}
where the object {\tt Pol} denotes a spin projection matrix $P_{\nu\mu}$, or more complicated substitutions including color contractions
\begin{verbatim}
In[5] :=
  QuarkContract[
  Eps[a, b, c] Eps[a', b', c'] Op[CI[{a, a'}], SI[{mu, nu}]]**
  Op'[CI[{b, b'}], SI[{mu, rho}]]
  ]
Out[5] = 
  quarkContract[{1, 3}, Op, Op'][CI[{c', c}], SI[{nu, rho}]]
\end{verbatim}
where the {\tt quarkContract[\#\#]} function is defined within
QDP++\footnote{Note that {\tt Op } and {\tt Op'} correspond to \emph{source1}
and \emph{source2} as given in the manual of QDP++, respectively.}.
Additionally we provide an {\tt Uncontract} function, which will split
contracted expressions into terms with open color and spin indices. Note that
this function will introduce a color and spin index irrespective of the
operator at hand, e.g.  a color neutral object has superfluous color indices
after the uncontract operation. We define $\delta$ functions to be independent of
their index type and it suffices to write {\tt DD[a,b]}$=\delta_{ab}$.
\section{Example applications}
Let us illustrate the capabilities of the package by means of two examples. 
\subsection{$\Lambda$ Baryon}
Suppose we are interested in the large time behavior of the matrix element
\begin{eqnarray}
\mathcal{M} = \langle \mathcal{O}_\mu \overline{\mathcal{O}_{\mu'}}\rangle P^{\mu'\mu}
\end{eqnarray}
e.g.  the two point function of a $\Lambda$  baryon.  Let us write the
interpolating operator for the $\Lambda$ baryon \cite{Gattringer:2010zz}
\begin{eqnarray}
	\mathcal{O}_\mu & = & \epsilon ^{abc} \Gamma^A_{\mu\alpha} \left( 2 s_a^\alpha (u_b^T \Gamma^B d_c) +
d_a^\alpha(u_b^T \Gamma^B s_c) - u_a^\alpha ( d^T_b \Gamma^B s_c)\right),\\
\overline{\mathcal{O}}_\mu & = & \epsilon ^{abc} \left( 2 (\bar u_a^T
\tilde\Gamma^B \bar d_b) \bar s _c^\alpha+ (\bar u_a^T \tilde \Gamma^B \bar s_b) \bar
d_c^\alpha - ( \bar d^T_a \tilde \Gamma^B \bar s_b) \bar u _c^\alpha \right)\Gamma^A_{\alpha\mu},
\end{eqnarray}
where 
$\mu$ and $\alpha$ denote spin indices with summation over repeated indices implied. 
Note that especially the matrices $(\Gamma^A,\Gamma^B)$
have spin indices, $A$ and $B$ are not summed over.
These operators in QCT read
\begin{verbatim}
In[6] :=
  OP =  
  Eps[a, b, c] ** (Gamma^A)[SI[{mu, alpha}]] ** (2 ** 
  Field[s, a, alpha, x] ** Field[u, b, beta, x] ** 
  (Gamma^B)[SI[{beta, gamma}]] ** Field[d, c, gamma, x] + 
  Field[d, a, alpha, x] ** Field[u, b, beta, x] ** 
  (Gamma^B)[SI[{beta, gamma}]] ** Field[s, c, gamma, x] - 
  Field[u, a, alpha, x] ** Field[d, b, beta, x] ** 
  (Gamma^B)[SI[{beta, gamma}]] ** Field[s, c, gamma, x])

In[7] :=
  OPBar =
  Eps[a', b', c'] ** (2 ** 
  FieldB[u, a', gamma', y] ** (Gamma^BT)[SI[{beta', gamma'}]] ** 
  FieldB[d, b', beta', y] ** FieldB[s, c', alpha', y] + 
  FieldB[u, a', gamma', y] ** (Gamma^BT)[SI[{beta', gamma'}]] ** 
  FieldB[s, b', beta', y] ** FieldB[d, c', alpha', y] - 
  FieldB[d, a', gamma', y] ** (Gamma^BT)[SI[{beta', gamma'}]] ** 
  FieldB[s, b', beta', y] ** FieldB[u, c', alpha', y]) ** 
  (Gamma^A)[SI[{alpha', mu'}]] 
\end{verbatim}
Within QCT the two point function can be computed as
\begin{verbatim}
In[8] := 
  Timing[Contracted = 
  WickContract[P[SI[{mu', mu}]] ** OP ** OPBar];]
Out[8] = 
  {0.016128000000000142, Null}
\end{verbatim}
where the result still has yet to be contracted spin and color indices. Note
that the computational time is negligible for a matrix element comprised of 6
quark operators.  Simplifying the expressions further
\begin{verbatim}
In[9] := 
  Result = QuarkContract[Contracted];
  ToQDP[
  Simplify[Result], {Gamma^B -> GammaB, Gamma^A -> 
  GammaA, Gamma^BT -> GammaBT}]
\end{verbatim}
gives the result
\begin{verbatim}
Out[9] =
/* 
Result is of type Scalar
{S^d(y,x) -> quarkProp1, S^s(y,x) -> quarkProp2, 
 S^u(y,x) -> quarkProp3}
*/ 
trace(quarkContract13(quarkProp1,GammaB*quarkProp2*GammaBT)*
transposeSpin(GammaA*P*GammaA*quarkProp3))
+
2*trace(quarkContract13(quarkProp3,GammaB*quarkProp1)*
transposeSpin(GammaA*P*GammaA*quarkProp2*GammaBT))
+
trace(quarkContract13(quarkProp3,GammaB*quarkProp2*GammaBT)*
transposeSpin(GammaA*P*GammaA*quarkProp1))
+
2*trace(quarkContract13(GammaB*quarkProp2,quarkProp1*GammaBT)*
transposeSpin(GammaA*P*GammaA*quarkProp3))
.
.
.
\end{verbatim}

The {\tt QuarkContract} function, as shown in the previous section, contracts
all open color and spin indices and identifies special patterns, e.g. resulting
in QDP++ {\tt quarkContract[\#\#]}, {\tt transposeSpin} $\dots$  function
calls.

\subsection{Sequential Source}
Another illustrative example is a  generic 3 point function
\begin{eqnarray}
\mathcal{M} = \langle \mathcal{O} \mathcal{J} \overline{\mathcal{O}} \rangle
\label{eq:three_pt}
\end{eqnarray}
Let us assume the following operators, e.g. nucleon interpolating operators and
a flavor preserving current, again adopting a vector notation,
\begin{eqnarray}
\mathcal{O} &=&  \epsilon^{abc} O^A u_a \left( u_b \Gamma^B d_c\right) ,\\
\overline{\mathcal{O}} &=& - \epsilon^{a'b'c'} \left( \bar d_a' \tilde\Gamma^B
\bar u_b'\right) \bar u_c' \tilde O^A ,\\
\mathcal{J} &=& \bar d J d .
\end{eqnarray}
In QCT the operators, amended with explicit spin and color indices, read
\begin{verbatim}
In[10] := 
  OP = Eps[a, b, c] ** OA[SI[{mu, alpha}]] ** 
  Field[u, a, alpha, y] ** Field[u, b, beta, y] ** 
  GammaB[SI[{beta, gamma}]] ** Field[d, c, gamma, y]

In[11] := 
  OPBar = -Eps[a', b', c'] ** FieldB[d, a', alpha', x] ** 
  GammaBT[SI[{alpha', beta'}]] ** FieldB[u, b', beta', x] **
  FieldB[u, c', nu', x] ** OAT[SI[{nu', mu'}]]

In[12] :=
  Jc = FieldB[d, f, sigma, z] ** J[sigma, rho] ** 
  Field[d, f, rho, z] 

In[13] := 
  MatrixElement = 
  OP ** Jc ** OPBar ** P[mu', mu]
\end{verbatim}

Now the matrix element can be parametrized as 
\begin{eqnarray}
\mathcal M = \Sigma \left(J S^{d}(z,x)\right)
\end{eqnarray}
where for the sake of readability we have omitted all volume sums and fourier
modes. The term $\Sigma$ is called the sequential propagator. The corresponding
source can be constructed using QCT, where one starts from the Wick contracted
matrix element, projects out the part proportional to $(JS^{d}(z,x))$ and
subsequently applies the Dirac equation to generate the sequential source - for
a discussion of the technique see e.g. \cite{Martinelli:1988rr}.

First to get the matrix element we write

\begin{verbatim}
In[14] :=
  MatrixElementContracted = 
  WickContract[MatrixElement] /. DE[__, {a_, a_}][__] -> 0
\end{verbatim}
where the replacement gets rid of disconnected pieces, i.e. propagators with
identical start and end points. The package implements a projection operator
{\tt DEProject} for propagators, i.e.
\begin{eqnarray}
S_{\nu_1,\nu_2}^{f,b_1, b_2}(x,z)  \text{Proj}_{f,\mu_1,\mu_2}^{a_1, a_2}(x,z) & = & \delta_{a_1 b_2} \delta_{a_2 b_1} \delta_{\mu_1 \nu_2} \delta_{\mu_2 \nu_1} \\
\mathcal{M}(S^f(x,z)) \text{Proj}_{\mu_1,\mu_2}^{f,a_1, a_2}(x,z) & =  & \mathcal{M}_{\mu_1,\mu_2}^{a_1, a_2}(x,z) 
\end{eqnarray}
Thus the easiest way to project out the sequential propagator is to
replace the current with a $\delta$-function and project out the
propagator connecting the source point x with the current insertion point z 
\begin{verbatim}
In[15] := 
  MatrixElementConnectedProj = 
  Expand[MatrixElementContracted] /. 
  J[sigma, rho] -> DD[rho, sigma];

In[16] :=
  SeqProp = 
  Contract[Expand[
  Uncontract[MatrixElementConnectedProj]
  DEProject[{d, d}, {x, z}][CI[{a1, a2}], SI[{mu1, mu2}]]]];
\end{verbatim}
The expression SeqProp is equivalent to $\Sigma_{\mu_1\mu_2}^{a_1 a_2}(x,z)$.
Next to find the source that generates this propagator we multiply with the
inverse Dirac propagator from the right, i.e.
\begin{eqnarray}
\eta_{\mu_1 \mu_3}^{a_1 a_3}(x,y) & = & \Sigma_{\mu_1 \mu_2}^{ a_1 a_2 }(x,z) 
S^{-1,a_2 a_3}_{\mu_2\mu_3}(z,y)
\end{eqnarray}

In QCT this reads 

\begin{verbatim}
In[17] := 
  QuarkContract[
  Uncontract[Contract[SeqProp]]*
  DEInverse[{d, d}, {y, z}][CI[{a2, a3}], SI[{mu2, mu3}]]]

Out[17] = 
  (quarkContract[{1, 4}, OAT . P . OA . DE[{u, u}, {x, y}] . 
  Transpose[GammaBT], DE[{u, u}, {x, y}]] . GammaB)[CI[{a1, a3}], 
  SI[{mu1, mu3}]] + 
  (transposeSpin[quarkContract[{3, 4}, DE[{u, u}, {x, y}] . 
  Transpose[GammaBT], OAT . P . OA . DE[{u, u}, {x, y}]]] . 
  GammaB)[CI[{a1, a3}], SI[{mu1, mu3}]]
\end{verbatim}
Note that to obtain the sequential source we applied the Dirac operator from the
right
\begin{eqnarray}
\sum_y \Psi(y)  S^{-1}(y,x) &=& \eta(x) \label{seqprop}
\end{eqnarray}
Usually lattice QCD codes implement a solver, where the solution from point x
to point y 
\begin{eqnarray}
\sum_y S^{-1}(x,y) \Psi(y) = \eta(x)
\end{eqnarray}
is calculated applying the Dirac operator from the left.\\
Rewriting Eq.(\ref{seqprop})
\begin{eqnarray}
\sum_y  S^{-1}(y,x)^\dagger \Psi(y) ^\dagger&=& \eta(x)^\dagger\\
\sum_y  S^{-1}(x,y) \gamma_5 \Psi(y) ^\dagger&=& \gamma_5\eta(x)^\dagger
\end{eqnarray}
one can however easily construct the sequential propagator using
\textit{ordinary} solves.

\subsection{Visualization}
The package allows for a visualization of the contractions, where the vertices are depicted as
circles and the propagators are drawn as lines. In QCT the
function {\tt GraphWC} expects the Wick contracted expression and returns
the distinct contractions, i.e. any prefactors are discarded, as graphics
objects. A basic elimination of identical graphs is performed, however the
output is not guaranteed to give the minimal set of contractions, where
further simplifications may be achieved, e.g. using reindexing.
The visualization of the contractions for the matrix element  Eq.~(\ref{eq:three_pt}) is performed via
\begin{verbatim}
In [18] :=
  GraphWC[MatrixElementContracted = WickContract[MatrixElement]]
\end{verbatim}
\begin{figure}[h]
\begin{center}
\includegraphics[width=.7\textwidth]{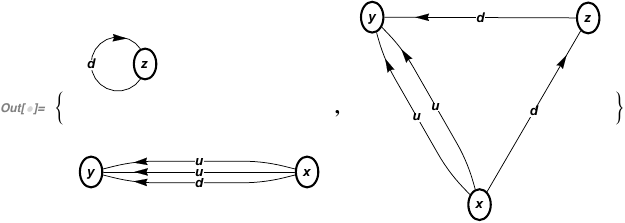}
\caption{Visualization of the contractions for the three point function of  Eq.~(\ref{eq:three_pt}).}
\label{fig:three_point}
\end{center}
\end{figure}
where the output is shown in Fig.~\ref{fig:three_point}. Internally {\tt GraphWC} uses the Mathematica builtin function {\tt GraphPlot} and therefor accepts the same options. Especially the options concerning vertex placement are the same, e.g.
\begin{verbatim}
In [19] :=
  GraphWC[WickContract[MatrixElement],
   VertexRules -> {x -> {-1, 0}, y -> {1, 0}, z -> {0, 1}}]
\end{verbatim}
which produces the output shown in Fig.~\ref{fig:three_point_vert}.
\begin{figure}[h]
\begin{center}
\includegraphics[width=.7\textwidth]{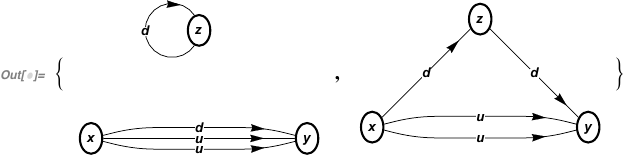}
\caption{Visualization of the contractions for the three point function (same as Fig.~\ref{fig:three_point}) using different vertex coordinates.}
\label{fig:three_point_vert}
\end{center}
\end{figure}
\section{Summary}
The aim of the package is to simplify the straightforward but error prone task
of computing matrix elements in QCD. To that end we implemented routines that
perform Wick contractions on quark operators, further simplify the results and
write out QDP++ expressions directly calculable within a C++ program. Since the
results are automatically generated, in most cases,  manual fine tuning will be
necessary to obtain the most efficient implementation of the matrix element at
hand. Nevertheless QCT is, at the very least, useful to check the correctness
of the implementation. The code is open source \cite{Djukanovic:2015dd}.
\section*{Acknowledgments}
The author would like to thank T. Harris for useful comments on the manuscript.

\end{document}